\documentstyle[12pt,aaspp4,flushrt]{article}

\begin{document}


\title{Infrared Observations of AR Ursae Majoris: \\
Modeling the Ellipsoidal Variations}

\author{Steve B. Howell}
\affil{Astrophysics Group, Planetary Science Institute \\
620 N. 6th Avenue, Tucson, AZ  85705}

\author{Dawn M. Gelino \& Thomas E. Harrison}
\affil{Department of Astronomy, \\
New Mexico State University, Las Cruces, NM 88003}

\begin{center}
Submitted to \sl{The Astronomical Journal}
\end{center}

\begin{abstract}
We have obtained time-series infrared photometry for the highly magnetic 
cataclysmic variable AR UMa.
Our J and K' band observations occurred during a low state and they show a 
distinctive
double-humped structure. Using detailed models for the expected ellipsoidal
variations in the infrared 
due to the non-spherical secondary star, we find that the most likely
value for the system inclination is 70$^o$. We also model low state V band 
photometry and find that its observed double-humped structure is {\it not} 
caused by ellipsoidal variations, as they have been ascribed to, but are
due to beamed cyclotron radiation. We use this result to estimate the 
magnetic field strength of the active southern accretion region (B$\la$ 
190 MG) and its magnetic longitude ($\psi_{S}$$\sim$330$^o$). 
\end{abstract}

{\it Subject headings:} cataclysmic variables
$-$ stars: binaries: close
$-$ stars: magnetic
$-$ stars: low mass

\newpage

\section{Introduction}

AR Ursae Majoris is the highest magnetic field strength polar known. Polars,
highly magnetic cataclysmic variables (CVs), are interacting binaries which 
contain a white dwarf primary and a low mass secondary star. The secondary 
fills its Roche lobe and transfers mass to the gravitational potential
of the more massive white dwarf. The white dwarf posseses a strong
(usually $\sim$10-60 MG) magnetic field which diverts the transfered material 
from its ballistic trajectory and funnels it along
magnetic flux tubes. This confinement of the accreted material results in
the formation of accretion columns which impact the white dwarf at one, 
or both, of its magnetic poles.

Detailed observational work on AR UMa has been performed by
Remillard et al. (1994), Schmidt et al. (1999), Szkody et al. (1999), and
references therein. These authors have discussed results of
visible, X-ray, and EUV observations of this binary including determinations 
of an orbital period of 115.9 minutes, a white dwarf mass of 0.6$\pm$0.2
M$_{\odot}$, and a binary inclination of 40$^o$ to 60$^o$. 
AR UMa, like all polars, shows modulations in its brightness over periods
of weeks to months which is believed to be caused by changes in the rate of 
mass accretion from the secondary star. The cause of these changes in \.M are 
not entirely understood, but may be related to star-spot activity on the 
secondary (Howell et al. 2000a). 
During high states, the active accretion region in AR UMa
is small in size but very hot ($\sim$ 250,000 K; Szkody et al. 
1999; Belle et al. 2000) and located near the south rotation pole of 
the primary star
(co-latitude of 10$^o$-35$^o$).
High state observations reveal V=15.1 mag and a dramatic 
change in EUV and X-ray emission, during which AR UMa becomes the brightest 
EUV source in the sky.
AR UMa, however, spends most of its time in a low 
state which is characterized by a V magnitude near 16.5 and weak high 
energy emission. 

Low state V and I band photometric observations of AR UMa show a double-humped
structure. The cause of these modulations is generally assumed  
to be ellipsoidal variations (Russell 1945). If true, infrared
observations of AR UMa would be expected to be 
dominated by ellipsoidal modulations and allow
detailed modeling of the secondary star. Such data could reveal quantitative
measures of the secondary star mass and temperature, as well as place additional
constraints on the system parameters. We therefore undertook IR photometric
observations of AR UMa in the J and K' bands.

In section 2, we describe our observations and data reduction, as well 
as present our IR photometric light curves. Section 3 describes our 
optical and infrared light curve modeling procedure using WD98, the newest 
version of the Wilson-Devinney light curve modeling code. We provide
details of how we chose the relavent input parameters and present the 
resulting models at V, I, J, and K'. Finally, Section 4 discusses the 
implications of these models.

\section{Observations \& Data Reduction}

AR UMa was observed using IRIM (see Joyce 1999) on the Kitt Peak 
National Observatory 2.1-m telescope on 13 and 14 February 2000.  
On 13 February, AR UMa was observed from 8:32 to 10:45 UT, and on 
14 February from 8:43 to 10:51 UT with each session covering
slightly more than one AR UMa orbital period. Photometric 
data were obtained in the IRIM J ($\lambda_c$=1.25 $\micron$) and 
K' ($\lambda_c$=2.15 $\micron$) filters. Our observing sequence 
consisted of two J images at one position, a beam switch, then two 
additional J images. We then switched to the K' filter, refocused,
and repeated the procedure. Each individual J image consisted of 
4 co-added frames of 10 seconds each, while the corresponding K' images
consisted of 10 co-added frames of 4 seconds each.

Both sky flats and dome flats were obtained.  Before processing, all of the
image data were linearized using the {\it irlincor} package within IRAF with 
the coefficients suggested for use in the IRIM User's Manual (Joyce 1999). 
After averaging the two images at one position, we subtracted them from
the average of the two images at the other position. These sky, and 
bias-subtracted images were then flat fielded using a flat constructed 
by averaging together a median of the dome and sky flats.

Aperture photometry was performed on AR UMa and two roughly equal brightness
nearby field stars (a close optical pair at RA=11:15:42, DEC=+42:55:20)
used as comparisons. Using the {\it phot} package in IRAF, instrumental 
magnitudes were measured and a differential light curve in both J and 
K' was generated with each point being the average of the two beam 
switched images. Mid-exposure times were heliocentric corrected. Proper 
magnitude uncertainties were determined according to the prescription 
given by Howell et al. (1988) with the average error being 0.05$\pm$0.02 
mag in both
bands.
A very few images were obtained through thick clouds and were not used 
in our light curve analysis. Since the conditions were non-photometric, 
standard stars were not observed on either night.  However, using the 
public release 2MASS data we determined a J and K' magnitude for a
random bright star (RA=11:15:50, DEC=+42:55:20) within our field of view.
Differential photometric results indicated that our two comparison 
stars did not vary by more than that expected due to photon statistics over 
the 
course of our observations. Therefore, we could use the bright 2MASS star
to assign standard J and K' magnitudes to our comparison stars. In turn, this
allowed us to transform our differential results for AR UMa into calibrated 
J and K' magnitudes. We found mean J and K' magnitudes for AR UMa on both 
nights of J=14.0$\pm$0.15 and K'=13.2$\pm$0.15. The 2MASS survey also 
measured AR UMa itself and found it to have J=14.1 and K'=13.3. These two 
measurements of J and K' for AR UMa are consistent within the 
uncertainies and within the
0.25 mag peak-to-peak variations in both J and K' light curves. Our resulting 
J and K' band light curves of AR UMa, phased to the Schmidt et al. (1999) 
ephemeris, are presented in Figure 1.

\section {Modeling the Light Curve of AR UMa}

\subsection{Observational Evidence}

The secondary star of AR UMa has been classified as an M6V$\pm$1 by 
Remillard et al. (1994), through analysis of an optical spectrum 
obtained in a low state. Using CV evolution models by Howell et al. 
(2000b), we can estimate the most likely mass for the secondary star 
in AR UMa. The secondary stars in CVs with orbital periods of 
less than two hours closely follow the lower main sequence mass-radius 
relation. Converting the determined mass (0.18 M$_{\odot}$) into an
assumed main sequence spectral type for the secondary star,
we would expect it to be an M4V with T$_{eff}$=3,150 K.
We note here that the spectral analysis by Schmidt et al. (1999)
gives a mass ratio of 0.3$\pm$0.1 which, when combined with their white dwarf 
mass 
(0.6 M$_{\odot}$), also yields a secondary star mass of 0.18$\pm$0.06 
M$_{\odot}$.

Schmidt et al. (1996) found that a white dwarf (log g = 8) model having
98\% of its projected area emitting at a temperature of T$_{eff} 
\sim$ 15,000 K, and only 2\% of the surface covered by an accretion spot 
of T$_{eff} \sim$ 35,000 K, fit a combination of IUE and optical spectra. 
Schmidt et al. further determined that the observed flux results in a 
WD radius of 8$\times$10$^8$ cm, consistent with a 0.6 M$_{\odot}$
white dwarf. These same authors concluded that the white dwarf and the 
secondary star each contribute roughly one half of the flux in the 
V-band during the low state.

Combining the mean low state V magnitude and the mean I band magnitude
from Remillard et al. (1994) with our mean J and K' 
magnitudes, we see that AR UMa has V-K'=3.3, I-K'=2.2, and  
J-K'=0.8. Within our errors, the infrared (I-K' and J-K') colors are 
consistent with an M4V secondary while the optical colors (V-I, V-J, or V-K'), 
however, are consistent with a K8V (Cox 2000; Glass 1999). Since a K8 star 
(0.60 M$_{\odot}$, appropriate for a CV with P$_{orb}$$\sim$ 5 h)
would not fit into the Roche lobe of a 1.9 h binary, we view these colors as 
caused by contamination by the hotter white dwarf. Another possibility for the
additional blue component may be the accretion region.  Szkody et al. (1999) 
used X-ray and EUV data to show that the accretion region temperature 
is quite high, near 250,000 K, during high accretion states. 
This temperature must be taken as an upper limit since, as noted above,
Schmidt et
al.. (1996) determined the accretion spot temperature 
to be substantially cooler only a few months
after a transition to a low accretion state.
This hot component might initially 
seem to be a likely candidate for the additional blue component
we need, but with its extremely small size (even at its hottest), 
it provides no meaningful
flux contribution ($<$1\%) to our light curves in either the 
infrared or in the optical.  This supplies further evidence of the 
white dwarf contamination in the optical.  

Light curves of magnetic cataclysmic variables, particularly those 
obtained during low states in the R or I visible bands or in the 
near infrared, are often observed to have a double-humped appearance 
with the cause of this shape generally assumed to be ellipsoidal 
variations due to the secondary star. Details of this claim are
rarely presented; the light curve shape is left to attest
to its cause simply by inference. Ellipsoidal variations should peak 
at orbital phase 0.25 and 0.75 (maximum projected area of the Roche 
lobe) and should have a cosine-like shape. Cyclotron emission radiated 
from near the accretion region can also provide increased light
at nearly these same phases. Near the white dwarf magnetic poles, 
the magnetic field lines are approximately
perpendicular to the surface and 
the location of the dynamically favored (active) pole is 
generally near binary phases 0.8-0.9. Thus, typical cyclotron 
beaming, strongest at 90$^o$ to the field lines (see Wickramasinghe 
and Ferrario 2000), will contribute increased system light preferentially 
centered near phases which are approximately coincident with the peak emission 
expected
from ellipsoidal variations. For example, 
G\"ansicke et al. (2000) show this to be the case at V for AM Herculis 
during a high state. Geometric factors such as the binary inclination, 
the accretion region co-latitude, and the location of the active magnetic pole,
as well as physical 
properties such as $B$ and \.M all conspire to make the exact phasing 
and shape of the photometric signal caused by cyclotron beaming
unique for any given polar. As we 
will see, even during polar low states, cyclotron beaming can provide 
a relatively strong photometric modulation, leading to a double-humped 
light curve similar in appearance to ellipsoidal variations.

Szkody et al. (1999) noted that the appearance of the AR UMa V band
light curve changes significantly between high and low states, showing 
a single peaked sine-like structure centered near phase 0.4 when bright, 
and a double-humped structure suggestive of ellipsoidal variations when 
faint.  We noted that the amplitude of the low state photometric humps 
seen at V (Szkody et al. 1999) was quite large and their flux peaked at 
orbital phases which were different from what was observed at I 
(Remillard et al. 1994; Schmidt et al. 1996) and in our infrared data.
Ciardi et al. (2000) show that during a high state of HU Aqr
the IR light curve is double humped but not well fit by ellipsoidal 
variations, their amplitude being too large and their 
shape being distinctively non-sinusoidal. However, low state IR 
photometry of this same star agrees with ellipsoidal models quite 
well (Ciardi et al. 1998). Recent IR spectral results for the polar 
ST LMi (Howell et al. 2000a) showed that during an extreme low state, 
the secondary star provides evidence for star-spot activity and
the IR light curve does not appear to be double-humped at all.

We did not obtain simultaneous V band measurements of AR UMa which 
could be used to confirm its accretion state during our IR observations.
However, K. Honeycutt (private communication) has informed us that
RoboScope observations provided relatively dense photometric monitoring of
AR UMa during the interval of 26 January to 6 March 2000, and the star 
was consistently between V=16.6 and V=16.8 mag, firmly in the low state.

Given the possible confusion related to the presence or lack of
ellipsoidal variations in polar light curves, particularly during 
low states, and the additional blue contribution seen in the colors of AR UMa,
we decided to attempt detailed modeling
of the low state light curves of AR UMa.

\subsection{Model Setup}

To model the light curves, we used WD98, the newest version of the 
Wilson-Devinney light curve program (Kallrath 1999; Wilson 1999).  
WD98 is an enhanced version of WD95 (Kallrath et al. 1998), updated 
with new features such as the addition of semi-transparent circumstellar 
clouds, a simple spectral line profile capability for fast-rotating 
stars, an option to work with either observed times or phases,
and conversion of all the variables to double precision.  Some
of the relevant features of WD98 include: Kurucz atmosphere models 
for numerous wavelengths, a choice of three different limb darkening 
laws, proximity and eclipse effects, the option for hot or cold
stellar spots, and several different modes of operation for various 
system geometries.  The program has been fully described in papers 
by Wilson and Devinney (1971) and by Wilson (1979, 1990, 1993).
A recent application of WD95 can be found in Milone et al. (2000). 

Briefly, WD98 works as follows. It takes the photospheres of the stars
and divides them up into a multitude of surface elements. The amount 
of light coming from each element is calculated based on the binary 
system input parameters.  All of these surface elements are then 
summed together, based on the line-of-sight geometry, to create the 
final light curve.

There are a large number of input parameters needed to generate a 
model light curve for a complex binary system such as AR UMa. We 
discuss each of the most important of these parameters in the 
following subsections.  We have made use of the best available 
literature system parameters and we list our wavelength independent 
input values to WD98 
in Table 1. Units are shown where appropriate.  We ran WD98 in a mode 
set up to produce a model for a semi-detached binary with the secondary
star automatically filling its Roche lobe.  We now discuss some of 
the input parameters in detail.

\subsection{Limb Darkening}

The most important parameter that affects both the shape and the 
amplitude of the ellipsoidal variations is limb darkening.  WD98 
allows you to choose from several different forms of limb darkening: 
linear, logarithmic, or square-root.  The linear law,
$$ I_{\lambda}(\mu) = I(1)(1 - x_{\lambda}(1 - \mu)), $$
was first investigated by Milne in 1921. In this equation, $I_{\lambda}$
is the beam intensity at wavelength $\lambda$, $\mu$ is the cosine 
of the angle between the atmosphere normal and the beam direction, 
and $x_{\lambda}$ is the limb darkening coefficient.  As an alternative 
to this, Klinglesmith \& Sobieski (1970) proposed the logarithmic law,
$$ I_{\lambda}(\mu) = I(1)(1 - x_{\lambda}(1 - \mu) - y_{\lambda} 
\mu {\rm ln}(\mu)), $$ where $y_{\lambda}$ is the non-linear limb-darkening 
coefficient. Di\'az-Cordov\'es \& Gim\`enez (1992) introduced the
square-root law,
$$ I_{\lambda}(\mu) = I(1)(1 - x_{\lambda}(1 - \mu) - y_{\lambda} (1 -
\sqrt{\mu})). $$ When fit to ATLAS atmosphere models, the logarithmic 
law appears to fit UV models the best, while the square-root law 
appears better at infrared wavelengths (Van Hamme 1993).  Models run 
by Claret (1998) for very low mass, solar metallicity stars (2000 K 
$\le$ T$_{eff}$ $\le$ 4000 K) indicate that the square-root law best
describes the intensity distribution in the infrared.  We ran test 
models of stars with equal temperature and gravity, and found that 
the logarithmic and square-root laws produced nearly indistinguishable 
light curves.  For the final models presented here, the square-root limb
darkening law was adopted.

As shown by Alencar \& Vaz (1999) the limb darkening coefficients of 
stars in close binaries can be effected by irradiation.  Szkody et al. 
(1999) attribute the high state V band light curve shape in AR UMa to 
irradiation of the secondary by the accretion region on the white 
dwarf surface. There is no evidence for such irradiation during low
states. In AR UMa, with a hot white dwarf primary and even hotter 
accretion region, it is important, however, to investigate whether 
irradiation has a perceivable effect on the limb darkening coefficients 
even during a low state. Due to the small size of the accretion spot 
and the low luminosity of the white dwarf, the models of Alencar \&
Vaz (1999) imply that the limb darkening coefficients used in our
AR UMa models will not be significantly affected by irradiation during 
a low state, and thus we have used the normal, non-irradiated,
coefficients.  Note, we are not ignoring the reflection effect due 
to irradiation here, we are merely stating that irradiation does not 
affect the limb darkening coefficients of AR UMa.

\subsection{Gravity Darkening}

The second most important static parameter that affects the amplitude
of the ellipsoidal variations is gravity darkening.  Gravity darkening 
(a.k.a. brightening) deals with the localized temperature of a star's 
surface.  The functional form for gravity darkening,
T$_{eff}$ $\propto$ g$^{\beta}$, is wavelength independent, and is 
not strongly affected by changes in either the mixing length or
composition of the star.  The amount of gravity darkening depends
on how energy is transported through the star, and is thus correlated 
with the mass of the star.  For low mass, convective stars, $\beta$=0.08, 
and for stars with radiative envelopes, $\beta \sim$1 (Lucy 1967).  
For the white dwarf primary in AR UMa, we have assumed no gravity 
darkening ($\beta$=0).  For the cool secondary stars used in our 
models, we have used a value of $\beta$=0.08.
                        
\subsection{Other input parameters}

There are a variety of other input parameters with less freedom in their
selection. For example, we must choose the values for the temperatures 
and monochromatic luminosities of the primary and secondary stars.  The 
monochromatic luminosities of the 15,000 K white dwarf proposed by Schmidt et 
al. (1996), were calculated from a simple blackbody model.  For the 
secondary stars, we calculated the corresponding luminosities using 
Bessell's tabulation (Bessell 1991) of absolute magnitudes of cool, late 
type, main sequence stars. For models with an M4V secondary, 
we used a secondary temperature of T$_{eff}$=3,150 K, and for models 
with an M6V secondary (as proposed by Remillard et al. 1994), we used a 
temperature of T$_{eff}$=2,800 K.  

In order to gauge what differences would occur between the two candidate 
secondary 
stars in AR UMa, we ran identical WD98 models for an M4V and an M6V star
using the same inclination angle of 70$^o$.
This choice of $i$ is not completely arbitrary as will be seen below.
It can be seen in Figure 2 that an M6V secondary would play a small role 
in the production of ellipsoidal variations at V, even at this high 
inclination.
Additionally, an M6V secondary star would not 
provide enough V flux to account for the 50\% contribution that has been 
proposed by Schmidt et al. (1996). 
While the two models are similar at J and K', we will see below that the 
I, J, and K' modulations are indeed ellipsoidal variations and therefore 
require the earlier type secondary. Based on the work presented in Howell 
et al. (2000b), and our discussion here related to Figure 2, we have 
chosen to use a M4V secondary star in all of our following models. 

The atmospheres of cool stars are fairly complicated and the details of 
their spectral energy distributions and any changes in such as a 
function of temperature, make for complex modeling (cf., Allard et al. 
1997). Stellar atmosphere codes have to take into account numerous atomic 
and molecular absorption features which affect the limb darkening 
coefficients.  In order to accurately model the limb darkening effects 
of the secondary star in AR UMa, we used a Kurucz model for its atmosphere,
the most complete atmosphere model available (Kurucz 1993; Kallrath \& 
Milone 1999).  Since the atmospheres of hot stars are much less complicated, 
particularly degenerate stars with thin atmospheres, we chose to model 
the white dwarf atmosphere as a blackbody.  

Irradiation of the secondary star atmosphere in close binaries such as AR 
UMa can be very important.  
WD98 calculates reflection/re-radiation of 
irradiation based on the bolometric albedos of the two stars.  The expected 
value for radiative envelopes like the white dwarf in AR UMa is unity.  
On the other hand, the bolometric albedo for the convective secondary star 
is expected to lie somewhere between 0.5 and 1, based on models run by 
Nordlund and Vaz (Nordlund \& Vaz 1990; Vaz \& Nordlund 1985).  This value 
is dependent on the amount of convection in the star: the smaller the mixing 
length parameter, $\alpha$ = $l/H_p$, the closer it is to a radiative 
atmosphere, and the higher the bolometric albedo.  Based on the average
of the albedos given in Table 3 of Nordlund \& Vaz (1990) for grey atmosphere 
models with a 6,000 K star irradiating a 4,500 K star, we chose to model 
AR UMa's secondary with a bolometric albedo of 0.676. The values published 
by Nordlund \& Vaz cover only a few binary star 
scenarios. We have chosen the one that most closely resembles our system.
Even though the system temperatures in Nordlund \& Vaz 
are not the same as those in AR UMa, the
scatter in the observed data is greater than the change in the light
curves with albedos differing by $\pm$0.1. 

WD98 has the capability to handle spots (hot or cold) on the component 
stars. In order to make our model more realistic, we added a small, 
circular bright spot (R$_{spot}$ = 0.05 R$_{WD}$; T$_{spot}$~$\sim$
250,000 K) to our white dwarf. This spot represents the active magnetic 
accretion region near the south pole of the white dwarf, and was modeled 
using parameters determined by Schmidt et al. (1999) and Szkody et al. 
(1999). Although the models were run with the accretion spot included, 
as noted earlier, the hot spot did not have any significant effect on
AR UMa's light curves due to its small projected area. Ciardi et 
al. (1998), included a similar accretion region in their spectral energy 
distribution models of the polar HU Aqr and also found that the hot accretion 
region contributed essentially 0\% to the flux in the visible and IR
spectral regions.
                                                           
Our AR UMa J and K' band observations and their corresponding WD98 
models for three possible values of the binary orbital inclination 
(50$^o$, 70$^o$, and 90$^o$), are presented in Figure 3.  The 
points represent the data while the lines represent the models.
We also ran WD98 models in the optical spectral region to compare with
V band data from Szkody et al. (1999) and maxima and minima I band 
data points taken from Remillard et al. (1994).  The V and I band 
models (Figure 4) were run with the same wavelength independent input 
parameters as the infrared models, but with appropriate monochromatic
luminosities, Kurucz model atmospheres, and limb darkening coefficients.
Our choice to model three different inclinations was inspired by the 
desire to see what differences inclination makes at these band-passes, to span
the range of possible system inclinations for AR UMa,  
and to understand how one might make use of such models to help determine system 
parameters.
         
\section{Results and Discussion}

Examination of the 50$^o$ model in Figs 3 \& 4 shows that the observed 
photometric amplitudes are not well matched by ellipsoidal variations
alone. At 50$^o$, an additional light source would be required in  all bands.
Increasing the inclination to 70$^o$, slightly out of the range proposed 
by Schmidt et al. (1999), we find a good correlation with the observations
except
at V. The inclination can not be much higher as
CVs with system inclinations larger then $\sim$72$^o$ will produce some 
observational evidence for an eclipse by the secondary star of either a 
part of the accretion stream, the white dwarf, or both. AR UMa shows no 
sign of such behavior. Forgetting the V band data for a moment, removal of the 
model
fits from the observed data 
shows that the observed J and K' modulations are well 
fit by ellipsoidal variations alone (Figure 5). From Fig. 4, we see 
that this would also hold true for I band observations. Our results 
clearly favor a binary inclination for AR UMa of near 70$^o$.

If we try to reconcile $i$=50$^o$, we note that
the J and K' light curves would show a 0.15 mag flux excess 
peaking near phases 0.25 and 0.75. These phases are coincident 
with those times during which the observer would view the gas stream
between the two stars at its maximum projected angle. Using an ad hoc gas 
stream model of rectangular dimensions and a uniform blackbody temperature 
of 6,500 K, we can only account for the 0.15 mag increase needed at J and K', 
if the emitting gas stream area was 50 times that of the white dwarf. This 
same model 
would have the gas stream producing 4 times the flux of the white dwarf
at V. Given the nature of the low state spectral observations (cf. Schmidt 
et al. 1999), this simple gas stream model seems unable to account for the 
excess light
needed if $i$=50$^o$. We could continue 
to modify our simplistic gas stream model by varying the temperature, 
emitting area, and even adding in a coupling region with some non-spherical 
shape, but the wisdom of Occam's razor\footnote{Occam's razor is ``not to 
compound hypothetical features or mechanisms of a scenario beyond necessity" 
({\it Entia non sunt multiplicanda praeter necessitatem}). However, William of 
Occam
(1285-1349) is only known to have written {\it Pluralitas non est ponenda sine
necessitas}, ``Plurality should not be posited without necessity" (Thorburn 1918). 
Noteable 
newer versions
of Occam's razor are: 
{\it We are to admit no more causes of natural things than such as are
both true and sufficient to explain their appearences} (Isaac Newton), {\it 
Everything
should be as simple as possible, but no simpler} (Albert Einstein), and 
{\it Keep it simple!}
(anonymous). The earliest quoted version of this razor is {\it Nature operates 
in the
shortest way possible}, attributed to Aristotle.} prevails.

For any value of $i$, it is clear that the
V data can not be fit by ellipsoidal variations alone. In Figure 5, we 
see that the ellipsoidal variation subtracted V light curve shows significant 
positive
residuals with peaks at
orbital phases 0.15 and 0.67, not at 0.25 and 0.75. Thus, we must seek an 
independent
source of these 0.1 mag photometric modulations.

We now explore 
cyclotron emission as a possible cause for the double-humped structure 
observed in the V light curve.
Schmidt et al. (1999) used circular polarization measurements to determine
that the magnetic longitude of the northern pole, $\psi_{N}$, 
was equal to 90$\pm$7 degrees, that 
is, at a right angle to the line of centers. Assuming diametrically opposed 
poles, the
southern active pole would be located near $\psi_{S}$ = 270$^o$  or binary 
phase 0.75. Observations of AR UMa made 
during a high state in the EUV spectral region (Szkody et al. 1999; Belle 
et al. 2000) show that the peak emission occurs near orbital phase 0.92. 
Sirk \& Howell (1998) modeled a number of similar EUV light curves for 
polars and found that in all cases, the value of $\psi_{active}$ is 
coincident with the phase 
of maximum flux. Using the EUV results for AR UMa, we estimate $\psi_{S}$ 
to be near longitude 330$^o$. 
This apparent change in $\psi_{S}$, compared with 270$^o$
implied by Schmidt et al. (1999), is not as dramatic as it first
appears, changing the physical location of the active region by very little
\footnote{ 
This is due to the fact that lines of longitude quickly become degenerate 
near a pole.}.  
If $\psi_{S}$=330$^o$ is correct (yielding $\psi_{N}$=150$^o$), 
then we would predict cyclotron emission to be 
a maximum near orbital phases 0.67 \& 0.17. These are 
the phases of 
peak emission observed in our residual V band light curve (Figure 5).

However, 
the estimated field strength of 230 MG for AR UMa places 
the fundamental cyclotron harmonic near 4660 \AA~ (for non-relativistic 
cyclotron radiation; Ingham, Brecher \& Wasserman 1976). With the 
fundamental and all higher harmonics (i.e., cyclotron humps) being blue-ward 
of the V band, it would seem unlikely that cyclotron radiation can provide 
the needed flux. 
Wickramasinghe \& Ferrario (2000) show that for almost all polars, the field 
strength 
of the two poles differ
by a ratio of 1.4-2, an effect assumed to be caused by a dipole offset 
from the white dwarf center by 
10-30\% of its radius. The active accreting pole, in all but one case, lies 
within 
$\pm$45$^o$ of the line of centers and in all cases the active pole is the 
weaker of the two.
The strength of the magnetic field (B=230 MG) in AR UMa was determined for the
northern pole and not the active accreting southern pole, thus the latter 
might be of weaker strength.

In order for 
our V band modulations to be caused by cyclotron emission, we would need the 
fundamental cyclotron harmonic ($\nu_c$) to move red-ward to at least 
within the V band.
Taking a red limit for $\nu_c$ at 5500 \AA, we require the magnetic field 
in AR UMa to be $\la$190 MG. Recent UV spectroscopy of AR UMa (G\"ansicke 2000)
has revealed
cyclotron humps and model fits estimate the magnetic field strength of the 
southern 
accreting pole
to be 160 MG. At 160 MG, the (southern) accreting pole is the weaker of 
the two and the ratio of the pole strengths in AR UMa is 1.44, placing 
it in the typical range for polars. To account for the observed
photometric amplitudes seen in V, the peak flux of the cyclotron emission
would need to be 1$\times$10$^{-16}$ ergs s$^{-1}$ cm$^{-2}$ \AA$^{-1}$~at 5500 
\AA.
Our upper limit for the southern pole field strength and our flux estimate for 
the
maximum needed cyclotron emission in V are both in agreement with the results 
provided
by G\"ansicke (2000).

In AR UMa, Schmidt et al. (1999) and Belle et al., (2000)
show that the active magnetic pole is 
the southern, dynamically favored one, and due to the combination of the 
system inclination and the location of this pole on the white dwarf, 
it is never in direct view for an Earthly observer. If the pole is not 
seen directly, the high state EUV 
observations provide information on flux emitted from above the accretion 
region, 
not directly from the white dwarf surface. The peak EUV emission at phase 0.92 
would then have to be interpreted as emanating from within the accretion column
itself, some small distance above the magnetic pole. 
It is indeed possible that in this very high field system, the magnetic
field close to the pole is not normal to the surface of the white dwarf 
and has a very complex structure. If true, the usual 
observational interpretations are invalid and without a direct view of the 
southern
active pole, its true location may be hard to determine.

The observed low state photometric light curves are consistent with
$i$=70$^o$, B$\la$190 MG, and $\psi_{S}$=330$^o$ given that I, J, and K' 
modulations are nearly 100\% ellipsoidal in nature and the V band modulations, 
centered at phases 0.15 and 0.67, are dominated by cyclotron emission. 
Our field strength of B$\la$190 MG for the southern accreting pole 
is in agreement with the recent value of 160 MG found
using model fits to cyclotron humps seen in UV spectra.
While our data are not of sufficient quality or 
time sampling to extend or refine the above arguments, one result that 
does come from 
this work is that any simple interpretation of a polar light curve as 
solely due to ellipsoidal variations should be viewed with caution. Each 
polar will provide a unique array of physical and geometric characteristics to 
deal 
with.

The authors wish to thank Kent Honeycutt for providing us with the RoboScope
data for AR UMa, and Paula Szkody for sending us the AR UMa V band data.
David Ciardi, Paula Szkody and Gary Schmidt provided valuable discussions 
related to this work. Boris G\"ansicke is thanked for providing his results 
prior to
publication. DMG would like to thank Josef Kallrath and Bob Wilson 
for the use of WD98, as well as their help in running and understanding the 
program. Comments from the anonymous referee were very useful and appreciated.
This paper made use of the on-line 2MASS database. 
SBH acknowledges partial support of this work by NSF grant AST 
9818770, NASA grant NAG5-8644, and an EUVE mini-grant. 
This research was also supported by a Grant-in-
Aid of Research from the National Academy of Sciences, through Sigma Xi,
The Scientific Research Society. DMG holds an American fellowship from the 
American Association of University Women Educational Foundation. KPNO/NOAO 
is operated by the Association of Universities for Research in Astronomy 
(AURA), Inc. under cooperative agreement with the National Science Foundation.

\begin{deluxetable}{lc}
\tablenum{1}
\tablecaption{Wavelength Independent WD98 Input Parameters for AR UMa}
\tablehead{
\colhead{Parameter}
 &\colhead{Value}
}
\startdata
Orbital Period\tablenotemark{a} (days) & 0.08050075 \\
Ephemeris\tablenotemark{a} (HJD phase 0.0) & 2450470.4309 \\
Semi-major axis (R$_{\odot}$) & 0.72 \\
Orbital eccentricity & 0.0 \\
Temperature of White Dwarf\tablenotemark{a} (K) & 15,000 \\
Temperature of M4V Secondary (K) & 3,150 \\                               
Mass ratio (M$_2$/M$_1$) & 0.3 \\
Atmosphere model (WD) & blackbody \\
Atmosphere model (M$_2$) &  Kurucz \\
Limb darkening Law & Square-root \\
Gravity darkening exponent (WD) & $\beta$=0.00 \\ 
Gravity darkening exponent (M$_2$) & $\beta$=0.08 \\ 
Bolometric Albedo (WD) & 1.000 \\
Bolometric Albedo (M$_2$) & 0.676 \\
\enddata
\tablenotetext{a} {From Schmidt et al., 1999}
\end{deluxetable}{}

\newpage
                                Figure Captions
                               
Figure 1.  AR UMa J band (top panel) and K' band (bottom panel) light curves. 
The data were obtained on 2000 February 13 and 14, with IRIM on the KPNO 
2.1-m telescope.  Here and throughout this paper we phase our heliocentric 
corrected data to the Schmidt et al. (1999) ephemeris. 

Figure 2.  AR UMa V band (V-C; Szkody et al., 1999, Figure 1) \& 
I band (Remillard et al.,
1994, Figure 7) maxima and minima data (points) are shown in the top 
panel. Our J \& K' band data (points) are shown in the bottom panel.  
The lines represent identical WD98 models for an 
M4V secondary star (solid line) and an M6V secondary star (dotted line), both
for a binary inclination of 70$^o$. The M4V and M6V models are similar at 
K' but clearly differ at the other wavelengths.

Figure 3.  AR UMa J band (top panel) and K' band (bottom panel) data (points)
from Figure 1, and three WD98 models.  The models were run with the input 
parameters listed 
in Table 1 for orbital inclination angles of: 50$^o$ (dotted line), 70$^o$ 
(solid line), and 90$^o$ (dashed line). 

Figure 4. (Top panel) AR UMa V band data (points) from Figure 1 of
Szkody et al. (1999). (Bottom Panel) Maximum and minimum I band data points
taken from Figure 7 of Remillard et al. (1994).  The lines represent WD98 
models for the same three orbital inclinations as in Figure 3 (50$^o$-dotted 
line, 70$^o$-solid line, and 90$^o$-dashed line). In this case, 
the 70$^o$ model fits the I band data very well, but none of the models 
provide an adequate fit at V.

Figure 5.  Differences (Observed AR UMa Data - Model) for the Szkody et al. 
(1999) V band data (top panel), our J band data (middle panel), and our K'
band data (bottom panel).  The differences are plotted for the 70$^o$ model 
and contain essentially no significant residuals at J and K'. However, 
the V data present 
a double-humped light curve of amplitude 0.1 mag with peaks at phases 0.15 
and 0.67. The V band modulation is almost entirely due to cyclotron beaming.

\newpage
\plotone{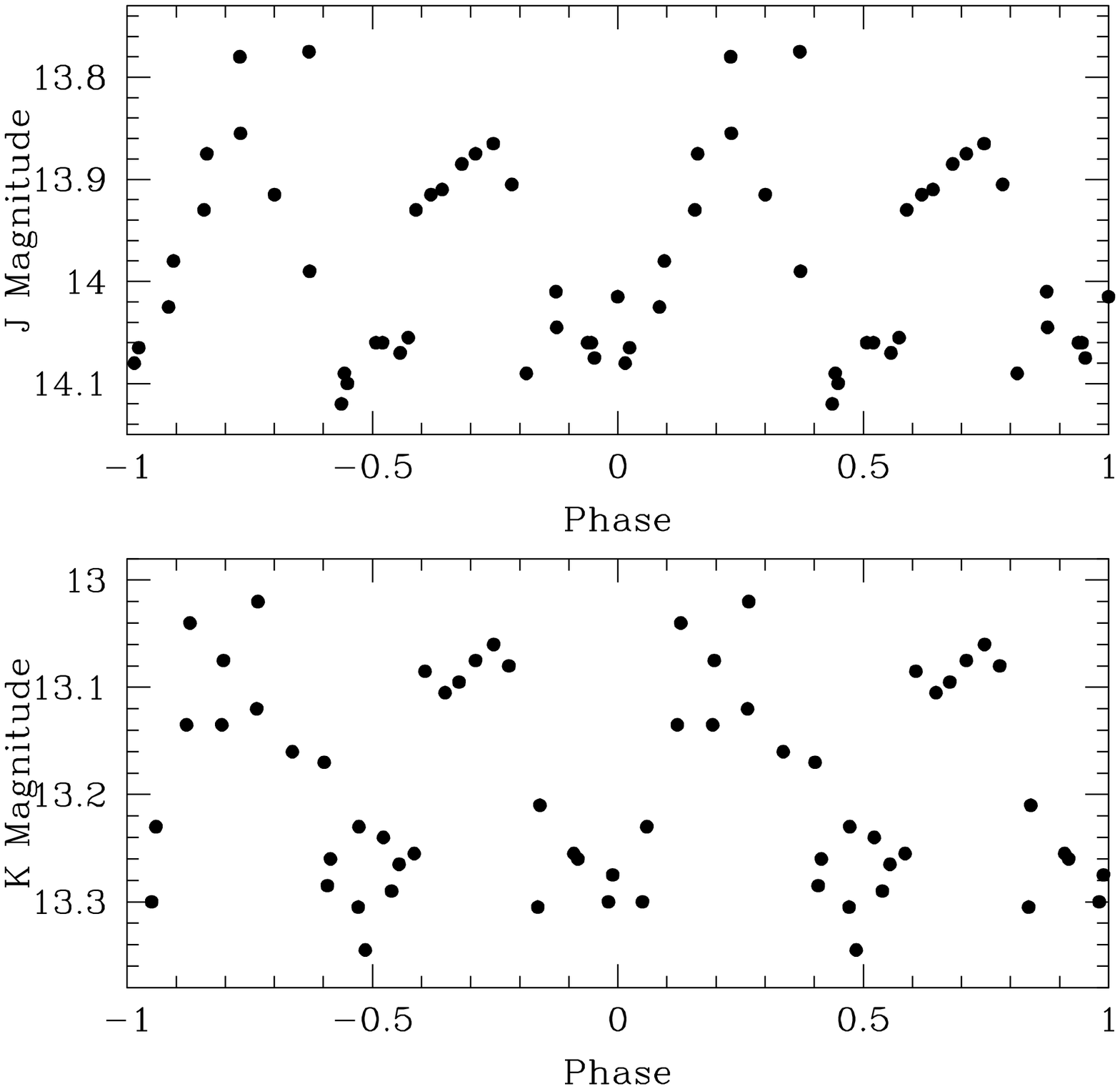}
\newpage
\plotone{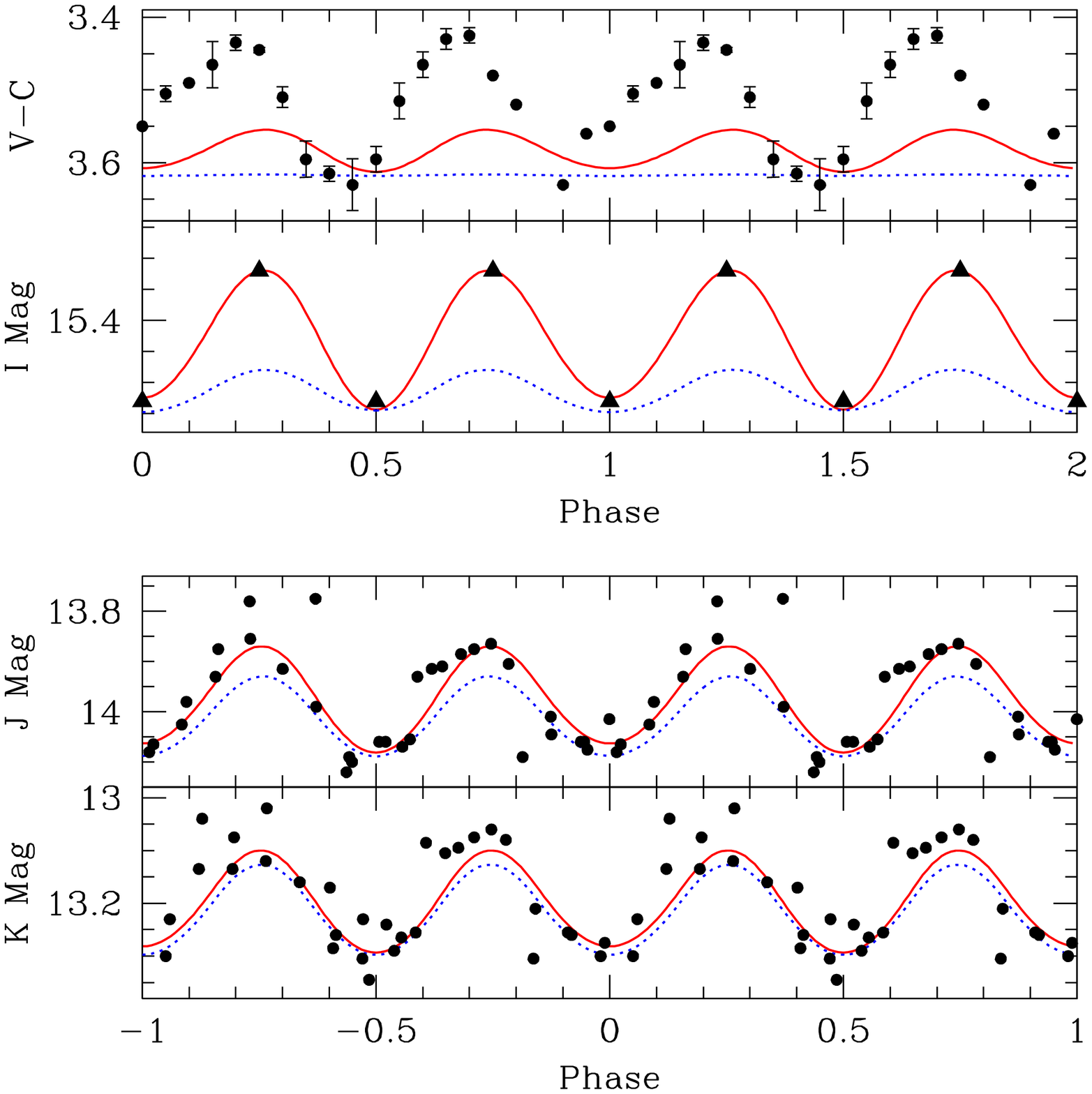}
\newpage
\plotone{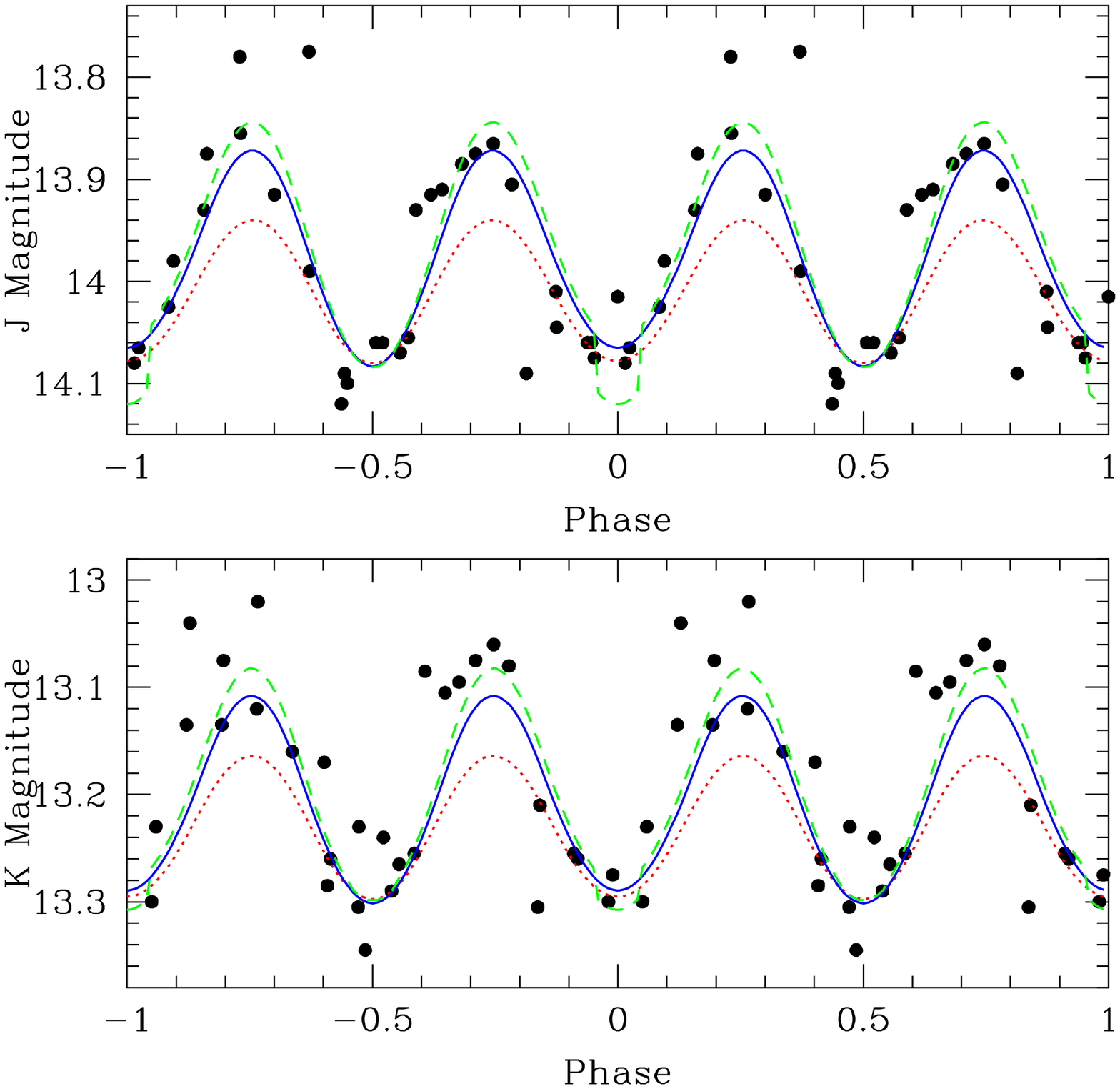}
\newpage
\plotone{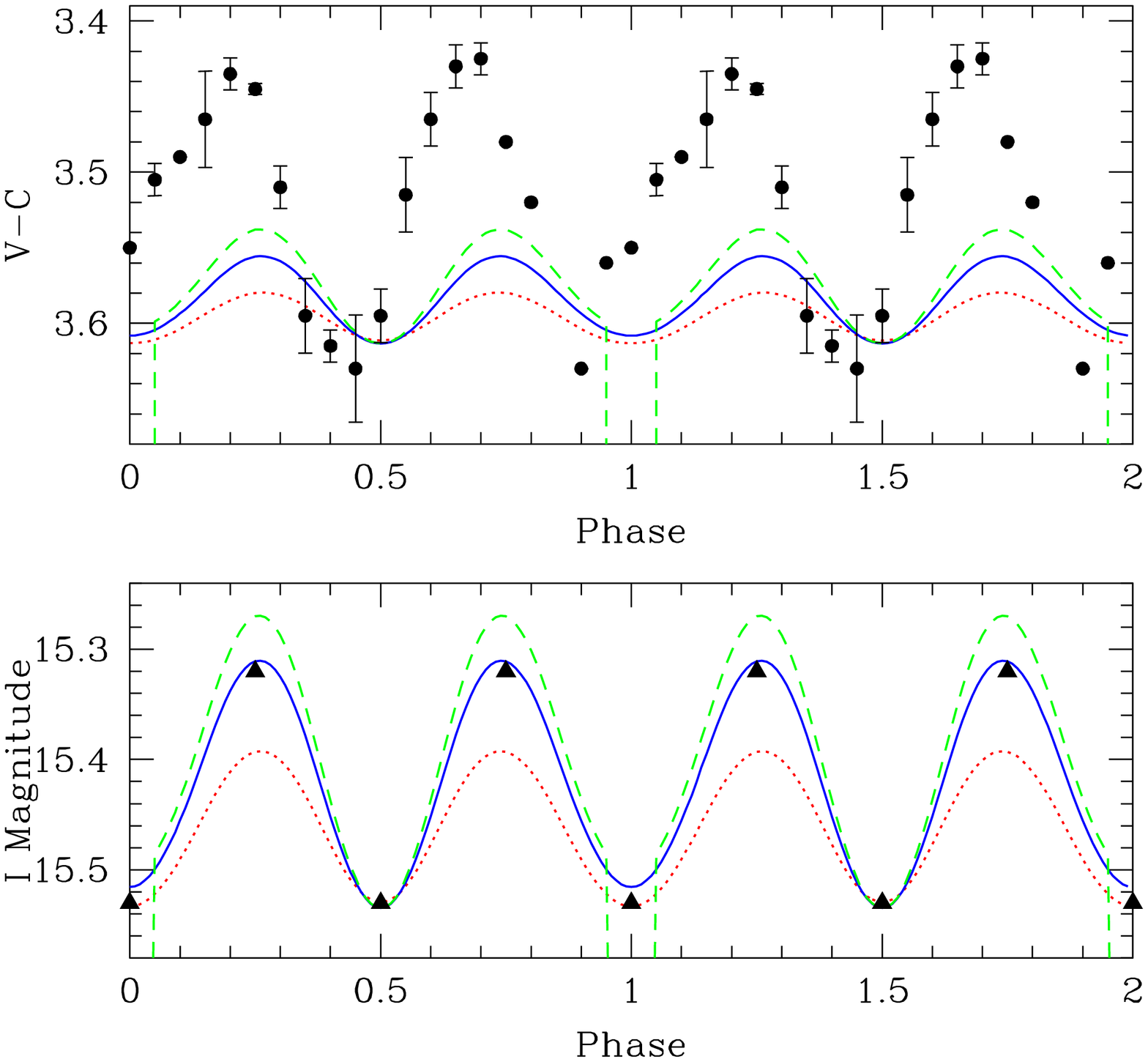}
\newpage
\plotone{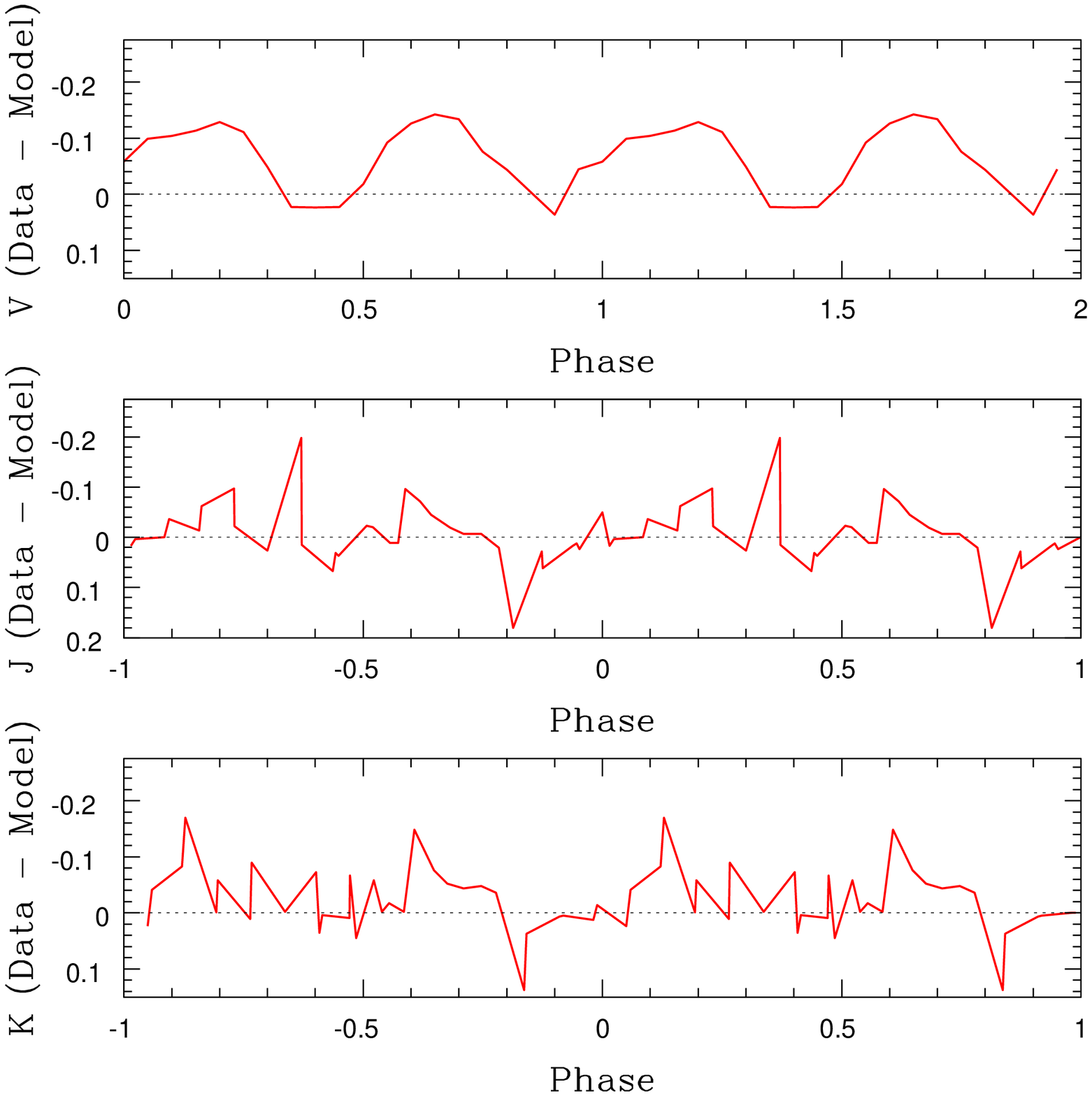}

\end{document}